\documentclass[seceq]{ptptex}

\usepackage{graphicx}

\title{
Quantum spin Hall phases%
}

\author{
Shuichi \textsc{Murakami}%
}

\inst{
Department of Physics, Tokyo Institute of Technology, 
2-12-1 Ookayama, Meguro-ku, Tokyo 152-8551, Japan\break
PRESTO, Japan Science and Technology Agency (JST), Saitama, Kawaguchi 
332-0012, Japan
}

\abst{
We review our recent theoretical 
works on the the quantum spin Hall effect.
First we compare edge states in various 2D systems, and 
see whether they are robust or fragile against perturbations. 
Through the comparisons we see the robust nature of edge states
in 2D quantum spin Hall phases. We see how it is protected by
the $Z_2$ topological number, and reveal the nature of the $Z_2$
topological number by studying the phase transition 
between the quantum spin Hall and insulator phases. 
We also review our theoretical proposal of the ultrathin bismuth 
film as a candidate to the 2D quantum spin Hall system.}

\begin{document}

\maketitle

\section{Introduction}
Recently spin Hall effect \cite{Murakami03a,Sinova04} 
has been studied theoretically and 
experimentally. It has shed new light onto the physics of the 
spin-orbit coupling.
In the spin Hall effect the spin-orbit coupling plays the role of the
spin-dependent effective magnetic field, causing the Hall effect
in a spin-dependent way.
As a related subject, quantum spin Hall (QSH) 
phase \cite{Kane05a,Kane05b,Bernevig06a}
is recently proposed theoretically. 
The QSH phase is a novel topological phase, where the 
bulk is gapped and insulating while there are
gapless states localized near the system boundaries.
This phase can be found among nonmagnetic insulators, and 
opened up a renewed interest onto nonmagnetic insulators.
This phase is a kind of topological order, and it is not
evident compared with other type of orderings such as magnetism
and superconductivity. 
This phase is not an ordered phase in the sense of Ginzburg-Landau
theory.
It is rather a topological order, which is encoded in the wavefunctions
themselves, and it appears only at the boundaries.
This topological order is hidden in the 
bulk, and appears as topologically protected gapless states at the
sample boundaries and interfaces.
The topological number, in the present case the $Z_2$ topological number
\cite{Kane05a},
characterizes whether the system is in the QSH phase or not.
In a sense, it plays the role of the ``order parameter''.
This phase can be realized  in 2D and in 3D \cite{Fu06b,Moore06,Roy};
the order guarantees the existence of gapless edge states for 2D and 
gapless surface states for 3D.

The QSH phase resembles the quantum Hall (QH) phase, while
there are several important differences. 
The QSH phase requires an absence of magnetic field, 
while the QH phase requires a rather strong magnetic field. 
Moreover, the QH phase is realized 
usually in 2D, and it is not easily realized in 3D because the
motion along the magnetic field usually does not become gapful.
In contrast, in the QSH phase there is no such built-in direction
and it can be easily realized in 3D as well as in 2D {\it without 
applying fields}.

These gapless edge states in 2D are peculiar
in the sense that it is robust against nonmagnetic disorder 
and interaction \cite{Wu06,Xu06a}. This is in strong constrast with usual
states localized at the boundary,  which are sensitive to 
the boundary roughness and impurities and so forth.

Since general readers may not be familiar with this issue,
in the first half of this paper we give a general 
instructive review for the whole subject: ``quantum spin Hall  
phase for pedestrians''. In the latter half we explain
our recent research on this topic. 
The paper is organized as follows. 
In Section 2 we consider edge states of various 2D systems 
and see whether they are robust or not. 
Surface states on the 3D QSH phases is also mentioned.
In Section 3 we see how the quantum phase transition between the 
QSH and insulator phases occurs. 
Section 4 is devoted to explanations of 
an existence of gapless helical edge states
based on the models obtained in Sec.~3.
In Section 5 we theoretically propose that the bismuth ultrathin film
will be a good candidate for the 2D QSH phase. In Section 6 we give concluding remarks.

\section{Edge States in Various Systems -- Fragile or Robust ?}

\subsection{Edge States in Graphene}
Graphene has been studied intensively in recent years.
One of the novel properties of graphene is the edge states.
It was theoretically proposed by Fujita et al.\ \cite{Fujita},
and was observed by STM experiments. This is a good 
starting point for studying the edge states in various systems. 

\begin{figure}
\centerline{\includegraphics[width=12cm]{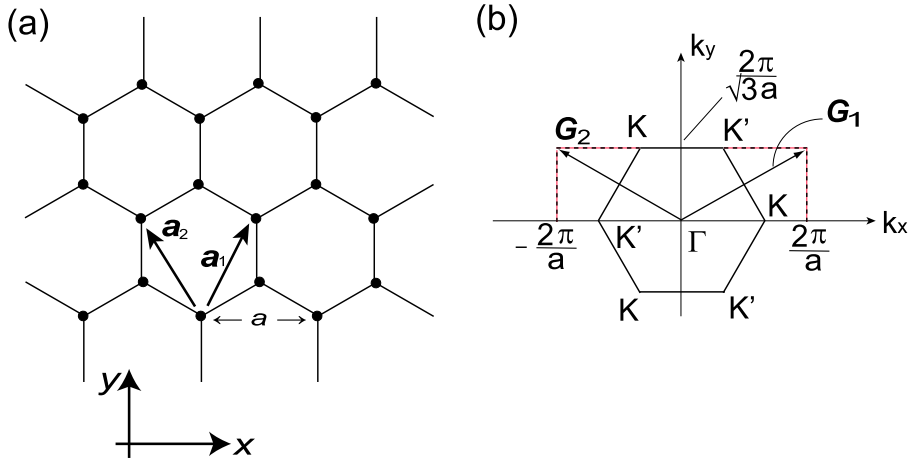}}
\caption{(a) Honeycomb lattice and (b) Brillouin zone corresponding to the honeycomb lattice.}
\label{fig:BZ}
\end{figure}

One simple way to see the edge states in graphene 
is the nearest-neighbor tight-binding model on 
a honeycomb lattice (Fig.~\ref{fig:BZ}(a)), described as
\begin{equation}
H=t\sum_{\langle i,j\rangle}c_{i}^{\dagger}c_j .
\end{equation}
We ignore other details of graphene, since they are inessential to 
the subsequent discussions. 
The primitive vectors are 
\begin{equation}
\mathbf{a}_{1}=\frac{a}{2}(1,\sqrt{3}),\
\mathbf{a}_{2}=\frac{a}{2}(-1,\sqrt{3}),
\end{equation}
and the reciprocal lattice vectors are
\begin{equation}
\mathbf{G}_{1}=\frac{2\pi}{a}\left(1,\frac{1}{\sqrt{3}}\right),\
\mathbf{G}_{2}=\frac{2\pi}{a}\left(-1,\frac{1}{\sqrt{3}}\right).
\end{equation}
The Brillouin zone is shown as Fig.~\ref{fig:BZ}(b). 

\begin{figure}
\centerline{\includegraphics[width=12cm]{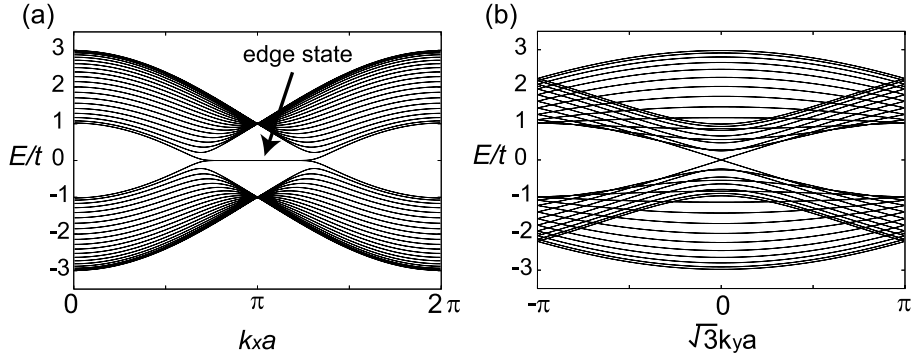}}
\caption{Band structure of the nearest-neighbor tight-binding model of the
the graphene 
in the ribbon geometry with (a) zigzag edges and (b) armchair edges.}
\label{fig:graphene}
\end{figure}

When we consider the tight-binding model on
an infinite system, the eigenenergies are
\begin{equation}
E({\bf k})=\pm t \left| e^{i{\bf k}\cdot {\bf a}_{1}}+
 e^{i{\bf k}\cdot {\bf a}_{2}}+1 \right|.
\end{equation}
The gap between the valence and the conduction bands
closes at two different points in the Brillouin zone, which are 
called $K$ and $K'$ points. Their  wavenumbers are
\begin{equation}
\mathbf{k}_{K}=\frac{1}{3}(\mathbf{G}_{1}
-\mathbf{G}_{2})=\left(\frac{4\pi}{3a},\ 0\right),\ 
\mathbf{k}_{K'}=-
\mathbf{k}_{K}.
\end{equation}
Around these points the dispersion is linear and forms
Dirac fermions.
To study edge states, we consider the system geometry with edges. For this purpose we make the system to 
have the ribbon geometry, having a finite width in one direction and
being infinite in the direction perpendicular to it. 
The honeycomb lattice allows
various types of edge shape, and we choose the 
zigzag edge and the armchair edge for example. The band structures 
for the two choices are shown in Fig.~\ref{fig:graphene} (a) and (b), 
respectively.
To understand the obtained band structures in this ribbon geometry,
we relate the bulk band structure and 
that of the ribbon as follows. 
In the ribbon geometry, the translational symmetry in one direction 
(perpendicular to the ribbon) is lost and the wavenumber along 
this direction is no longer a good quantum number. 
Therefore the bulk band structure is projected along this  
direction. This almost corresponds to the band structure 
for the ribbon geometry, calculated in Fig.~\ref{fig:graphene}(a) and (b). 
In the zigzag-edge case 
(Fig.~\ref{fig:graphene}(a)), however, 
the states located at $E=0$, $\frac{2\pi}{3}<k_x a< \frac{4\pi}{3}$
is outside the bulk-band projection.
This is nothing but the edge states, because it is located
in the bulk band gap, i.e. within the region where the bulk states 
are prohibited. 

We note that the above tight-binding model is with drastic 
simplification. In reality there are many factors which have
been ignored here. We thus ask ourselves whether the
above properties survive perturbations. 
For example, 
within the above tight-binding model, we can introduce various perturbations
which do not support edge states. 
When we vary the boundary conditions, the 
edge states may vanish. 
For example, the armchair edge does not support edge states. 
Another perturbation which kills edge states around $E=0$ is a staggered on-site potential,
although it may not be realized in real systems.

This fragility of the edge states arises because the edge states in 
graphene are not topologically protected, due to vanishing of the 
bulk gap.
In the following sections we see various kinds of 
robust edge states. All of these robust edge states
are associated with respective topological numbers. 
A more physical reason why the edge states in graphene 
are fragile is
that no current of any kind is flowing along the edge. 
As we see from Fig.~\ref{fig:graphene}(a), the edge states form a flat band, 
which means that 
the velocity along the edge is zero. 
The modes are localized 
and does not move along the edge. 
In fact, this flatness of the edge-state band is not a universal property;
some perturbations give a dispersion to the otherwise flat band.
For example, the next-nearest-neighbor hopping brings about a downward 
shift of the dispersion around $k_x=\pi/a$, compared with 
Fig.~\ref{fig:graphene} (a).
Even in this case, the net current along the edge sums up to zero.

\subsection{Edge States in Quantum Hall Systems}
In this section we consider an example of robust edge states. 
We consider the integer QH systems. 
The QH systems are realized experimentally in a two-dimensional
electron system in a strong magnetic field. 
In the QH systems the bulk is gapped while the edge has gapless 
edge states, which carries chiral current.
The number of chiral edge states, $\nu$, is a topological quantity, 
which does not change under weak perturbation.
In this case the robustness of the edge states comes from the 
topological number $\nu$. The bulk states bear a topological order,
described by the topological number $\nu$ called the Chern number.

The edge states of the QH system look very different from those
in the graphene discussed in the previous section. 
Nevertheless, by using a simple model we can relate them and 
discuss their differences.
It is the model proposed by Haldane \cite{Haldane88}.
It is a tight-binding model on the honeycomb lattice,
where the nearest-neighbor hopping is real, and the 
next-nearest neighbor hopping is complex. 
The Hamiltonian is given by 
\begin{equation}
H_{{\rm Haldane}}
=t_1\sum_{\langle i,j\rangle}c_{i}^{\dagger}c_j
+t_2\sum_{\langle\langle i,j\rangle \rangle}e^{-i\nu_{ij}\phi}c_{i}^{\dagger}
c_{j}+M\sum_{i}\xi_ic_i^{\dagger}c_i.
\label{eq:Haldane}
\end{equation}
Here $\nu_{ij}={\rm sgn}(\hat{{\bf d}}_{1}\times\hat{{\bf d}}_{2})_{z}=\pm 1$, 
where $\hat{d}_{1}$ and $\hat{d}_{2}$ are unit vectors along
the two bonds, which constitute the next-nearest neighbor hopping. 
$\xi_i$ represents a staggered on-site potential,
and takes the values $\pm 1$ depending on the $i$-th sites being
in the A or B sublattices, respectively.
This simply means that the next nearest neighbor 
hopping, going around the hexagonal plaquette in 
the clockwise (counterclockwise) way, obtains the phase $e^{i\phi}$
($e^{-i\phi}$).

The quantization of $\sigma_{xy}$ for this model without impurities
can be seen from the Kubo formula, and the resulting 
$\sigma_{xy}$ is rewritten as $\sigma_{xy}=\nu e^2/h$ \cite{Thouless82,Kohmoto85}, 
where 
\begin{equation}
\nu=\int_{{\rm BZ}}\sum_{n:{\rm filled}}
\frac{d^{2}\bf k}{2\pi} B_z^{(n)}({\bf k}),
\label{eq:nu}
\end{equation}
and 
\begin{equation}
B_z^{(n)}({\bf k})=
\frac{\partial A_y^{(n)}({\bf k})}{\partial k_x}-
\frac{\partial A_x^{(n)}({\bf k})}{\partial k_y}, 
A_i^{(n)}({\bf k})=-i\left\langle u_{n{\bf k}}\right|
\frac{\partial}{\partial k_{i}}\left|u_{n{\bf k}}\right\rangle.
\label{eq:Ch}
\end{equation}
The sum in Eq.~(\ref{eq:nu}) is taken over the 
filled bands. The quantity $\nu$ has novel properties arising from
topology. 
A na{\"i}ve application of the Stokes theorem casts 
Eq.~(\ref{eq:nu}) into a contour integral along 
the border of the Brillouin zone (BZ), and the periodicity of
the BZ results in $\nu=0$; however it is not true \cite{Kohmoto85}.
In some cases as the present one, 
the Bloch wavefunction cannot be expressed as a single continuous
function over the whole BZ; the BZ should be divided into 
pieces, on each of which the Bloch wavefunction is continuous.
At the borders between two ``pieces'' the Bloch wavefunctions
differ by a U(1) phase. The number $\nu$ is expressed in terms
of this phase difference. As a result one can show that
this quantity is quantized to be an integer, 
and it turns out to be the topological number called the Chern number 
\cite{Thouless82,Kohmoto85}.

This Chern number represents the number of chiral gapless edge states
going around the sample edge. 
Namely the existence of the gapless 
edge states is guaranteed by the Chern number.
This comes from the Laughlin argument \cite{Laughlin81}; 
we roll the two-dimensional
system into an open cylinder by attaching two edges on the opposite sides, and 
let a flux $\Phi$ penetrate the hole. The 
flux $\Phi$ 
is to be increased from 0 to a flux quantum. Then 
the number of electrons carried from one edge of the cylinder to the other is 
equal to the Chern number. These carried electrons are on the gapless 
edge states. Thus we can establish the correspondence between the 
number of chiral edge states and the Chern number.

To calculate the value of $\nu$ for the Haldane's model, it is not easy to 
calculate Eq.(\ref{eq:Ch}) directly, and analytical 
calculation looks almost impossible. 
Instead by considering the change of $\nu$ by a change of some parameter, 
we can easily calculate
$\nu$. 
First we notice that for $\phi=0$ the system is time-reversal 
symmetric, which yields $\nu=0$, because $\nu$ changes sign 
under time-reversal.
As long as the gap remains open, the Chern number 
$\nu$ is quantized to be an integer 
and 
cannot change when a parameter in the Hamiltonian is 
continuously changed. In the Haldane's model, 
by changing the system parameters, the gap closes only at
$K$ and $K'$ points.
At the gap closing the spectrum of Eq.~(\ref{eq:Haldane}) 
becomes linear in ${\bf k}$ near the gap-closing point, 
namely the massless Dirac fermion is formed at the gap closing.

\begin{figure}
\centerline{\includegraphics[width=8cm]{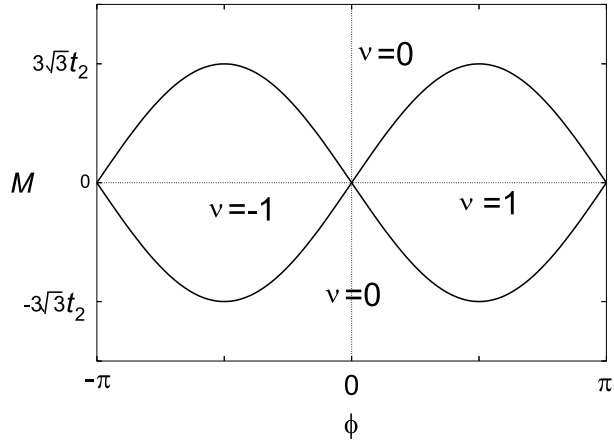}}
\caption{Phase diagram of the Haldane's honeycomb-lattice model.
$\nu$ represents the Chern number}
\label{fig:haldane}
\end{figure}

It is straightforward to show the following statement for the 
change of the Chern number at the gap closing.
We consider a $2\times 2$ Hamiltonian matrix $H(M,{\bf k})$, which 
depends on a parameter $M$.
Suppose the gap between the two bands closes at $M=M^{(0)}$ and
${\bf k}={\bf k}^{(0)}$. Then we have 
$H(M^{(0)},{\bf k}^{(0)})=E_{0}(M^{(0)},{\bf k}^{(0)})\hat{1}$, where 
$\hat{1}$ is an identity matrix, and we
can write
\begin{equation}
H(M,{\bf k})=E_{0}(M,{\bf k})\hat{1}+\sum_{i}a_{i}(M,{\bf k})
s_{i},
\end{equation}
where $s_{i}$ are Pauli matrices.
For notational brevity and convenience we write 
$k_0\equiv M$, $k_1=k_x$, $k_2=k_y$.
The coefficients $a_{i}$ is expanded in $M$ and ${\bf k}$ in  
the linear order as
\begin{equation}
a_{i}=\sum_{j=0,1,2}(k_j-k_j^{(0)})a_{ij}.
\end{equation}
Then one can show 
that the change of the Chern number $\nu$ across $M=M^{(0)}$ is
\begin{equation}
\nu(M=M^{(0)}+\delta)-\nu(M=M^{(0)}-\delta)
={\rm sgn}({\rm det}a).
\end{equation}
where $a$ is the matrix with elements $a_{ij}$ and $\delta$ is an 
infinitesimal positive number. 
Thus the Chern number changes by one at the gap closing. 
In the present case, the gap closes when $M=\mp 
3\sqrt{3}t_2 \sin\phi$,
with ${\bf k}=\pm {\bf k}_{K}=({\bf k}_{K},\ {\bf k}_{K'}$),
respectively. 
For the respective cases, the matrix $a$ is calculated by linearlizing
the Hamiltonian in the vicinity of the gap closing as
\begin{equation}
a=\left(\begin{array}{ccc}
1&&\\& \mp\sqrt{3}a/2 &\\ &&-\sqrt{3}a/2 
\end{array}\right)
\end{equation}
and ${\rm sgn}({\rm det}a)=\pm 1$. Thus when we increase $M$ across the 
value $\mp 
3\sqrt{3}t_2 \sin\phi$, the Chern number increases by $\pm 1$. 
From this we can easily elaborate the phase diagram as shown in 
Fig.~\ref{fig:haldane}.
As we have seen, for analytical calculation 
it is much easier to calculate the {\it change} of the 
Chern number, than to calculate the Chern number itself. 

What is the fundamental differnce between the QH phase and the grephene?
The difference comes from the bulk gap. The graphene 
is gapless while the QH phase has a bulk gap. 
The topological number in the QH phase protects the 
existence of the gapless edge states.

\subsection{Edge States in 2D Quantum Spin Hall Systems}
The QSH system is insulating in the bulk, while 
there are gapless edge states carrying a spin current. 
The simplest case of the QSH system is realized by superposing
two QH subsystems with opposite spins \cite{Bernevig06a}. 
We consider the up-spin subsystem with a QH state 
($\sigma_{xy}=e^2/h$), and the down-spin subsystem with 
a QH state ($\sigma_{xy}=-e^2/h$). The superposition of these states
is the QSH state. The edge states consist of two states
with opposite spins and velocities. These states 
are thus carrying a spin current. 
To realize this state we need a spin-dependent magnetic field, which 
can be produced by the spin-orbit coupling. 
This example is just a superposition of two QH systems with conserved
 spin $s^z$.
Nevertheless, in real systems, the spin-orbit coupling 
does not necessarily conserve $s^z$, and the above
two QH subsystems are mixed with each other. 
The next question is what happens when $s^z$ is no
longer a good quantum number. 
We consider this question in the following, and 
we see that the physics coming from topology survives
partially. 

This phase is realized in the model proposed by Kane and Mele 
\cite{Kane05a,Kane05b}. 
The model is given as
\begin{equation}
H_{KM}=t\sum_{\langle i,j\rangle}c_{i}^{\dagger}c_j
+i\lambda_{{\rm SO}}
\sum_{\langle\langle i,j\rangle \rangle}\nu_{ij}c_{i}^{\dagger}s_z
c_{j}
+i\lambda_{R}\sum_{\langle i,j\rangle}c_{i}^{\dagger}({\bf s}\times
\hat{{\bf d}}_{ij})_z c_j+ \lambda_{v}\sum_i \xi_{i}c_{i}^{\dagger}
c_{i}.
\end{equation}
$\xi_{i}$ represents a staggered on-site potential,
taking values $\pm 1$ depending on the sublattice index. 
$\hat{{\bf d}}_{ij}$ is the unit vector along
 the nearest neighbor bond from $i$ to $j$.
For a special case, if $\lambda_R=0$ and $\lambda_v=0$, the model 
conserves $s^z$ and it reduces to a superposition of two Haldane models:
\begin{equation}
H_{KM}(\lambda_R=0,\lambda_v=0)=H^{\uparrow}_{{\rm Haldane}}(\phi=-\pi/2)+
H^{\downarrow}_{{\rm Haldane}}(\phi=\pi/2).
\end{equation}

For generic cases where $s^z$ is not conserved,
the key ingredient is the time-reversal 
symmetry, which gives rise to the Kramers degeneracy 
between ${\bf k}$ and $-{\bf k}$. 
In the theory the wavenumbers satisfying ${\bf k}\equiv -{\bf k}$
$({\rm mod}\ {\bf G})$ play an important role. 
Such momenta are called the time-reversal-invariant momenta (TRIM),
and are expressed as ${\bf k}={\bf \Gamma}_i$ where ${\bf \Gamma}_{i=(n_1 n_2)}
=\frac{1}{2}(n_1{\bf G}_{1}+n_2{\bf G}_{2})$ with $n_1,n_2=0,1$ in 2D 
\cite{Kane05a,Fu06a,Fu07b}.
The $Z_2$ topological number $\nu$ is defined in the following way
\cite{Fu06a,Fu07b}. 
First we define a $(2N)\times (2N)$ matrix $w$, defined as
\begin{equation}
w_{mn}({\bf k})=\langle u_{-{\bf k},m}|\Theta
|u_{{\bf k},m}\rangle
\end{equation}
where $\Theta$ is the time-reversal operator
represented as $\Theta=i\sigma_y K$ 
with $K$ being complex conjugation.
$u_{m,\mathbf{k}}$ is the periodic part of the $m$-th Bloch wavefunction lying 
below $E_F$, 
and 
$N$ is the number of Kramers pairs below $E_F$.
This matrix $w({\bf k})$ is unitary at any ${\bf k}$,
and is also antisymmetric at ${\bf k}={\bf \Gamma}_{i}$.
Then for each TRIM we define an index $\delta_i$ as
\begin{equation}
\delta_{i}\equiv \frac{\sqrt{{\rm det}w({\bf \Gamma}_i)}}{
{\rm Pf}w({\bf \Gamma}_i)}.
\label{eq:Iasym}
\end{equation}
for ${\cal I}$-asymmetric systems \cite{Fu06a}, and 
\begin{equation}
\delta_{i}\equiv \prod_{m=1}^{N}
\xi_{2m}({\bf \Gamma}_i),
\label{eq:Isym}
\end{equation}
for ${\cal I}$-symmetric systems \cite{Fu07b}, where 
$\xi_{2m}({\bf \Gamma}_i)$ ($=\pm 1$) is the parity eigenvalue of the 
Kramers pairs at ${\bf \Gamma}_i$.
The index $\delta_i$ takes the values $\pm 1$. 
The $Z_2$ topological number $\nu$ is then defined as
\begin{equation}
(-1)^{\nu}=\prod_{i=1}^{4}\delta_{i}, \ \
\label{eq:Z2-2D}
\end{equation}
where the product is taken over the TRIMs ${\bf k}={\bf \Gamma}_{i}$.
Hence $(-1)^{\nu}$ takes only two values $\pm 1$, which means
there are only two distinct cases $\nu=\mathrm{even}$ and 
$\nu=\mathrm{odd}$. For simplicity of notations we henceforth call
these cases as $\nu=0$ and $\nu=1$.
If the resulting value is $\nu=1$ it is the QSH phase, and if 
it is $\nu=0$ it is the ordinary insulator. 
The expression of the $Z_2$ topological number is different between 
systems with and without ${\cal I}$-symmetry.
This shows a crucial role of the ${\cal I}$-symmetry in the 
theory of the QSH systems. It is because the ${\cal I}$ symmetry
is the only symmetry operation which relates between ${\bf k}$ and
$-{\bf k}$. 

\begin{figure}
\centerline{\includegraphics[width=12cm]{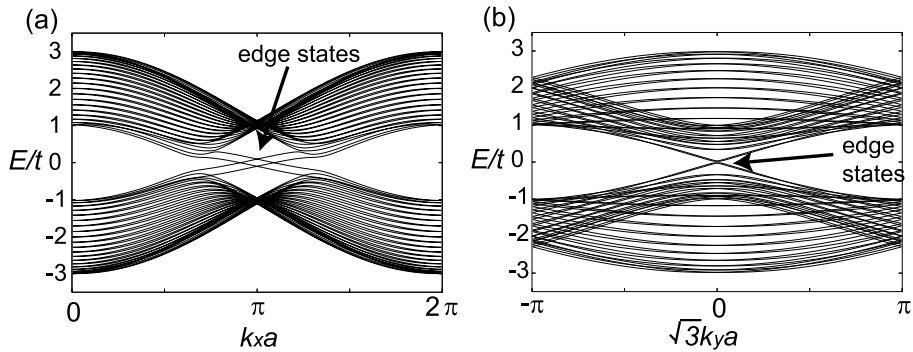}}
\caption{Band structure of the Kane-Mele model in the QSH phase
in the ribbon geometry with (a) zigzag edges and (b) armchair edges.
The parameters are $\lambda_v=0.1t$, 
 $\lambda_{SO}=0.06t$ and $\lambda_R=0.05t$.}
\label{fig:QSH}\end{figure}

\begin{figure}
\centerline{\includegraphics[width=12cm]{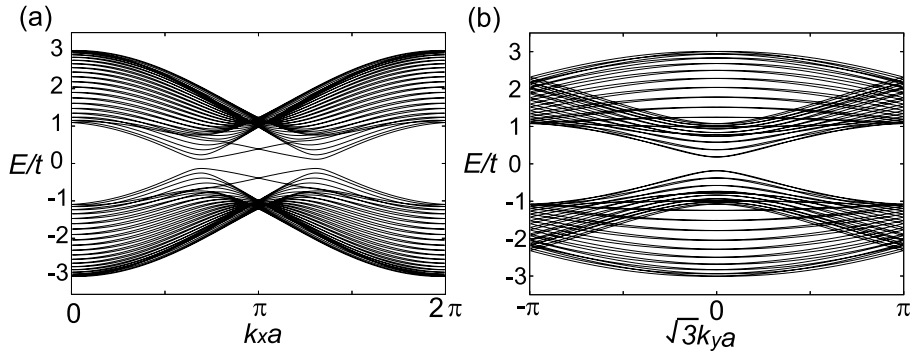}}
\caption{Band structure of the Kane-Mele model in the I phase
in the ribbon geometry with (a) zigzag edges and (b) armchair edges.
The parameters are $\lambda_v=0.4t$,
 $\lambda_{SO}=0.06t$ and $\lambda_R=0.05t$.}
\label{fig:SHI}
\end{figure}

To illustrate the physics of the $Z_2$ topological number, we calculate the 
band structure for geometries with edges. Let us consider
a ribbon geometry, which is finite in one direction and is infinite
in the other direction. 
To see the difference between two phases, QSH and ordinary insulator (I), 
we take up the sets of the parameter values for the respective 
phases, employed in Ref.~\citen{Kane05b}. The result
is shown in Figs.~\ref{fig:QSH} and \ref{fig:SHI} for the QSH and I
phases, respectively.
As we see from Fig.~\ref{fig:QSH}, 
there exist gapless edge states in the QSH phase, 
irrespective of the geometry.
In contrast, for the insulator phase
(Fig.~\ref{fig:SHI}) there are no gapless edge states. 
In fact there are edge states, but they do not
go across the gap. These edge states may or may not
cross the Fermi energy. Even if they cross the Fermi
energy, it is not an intrinsic property, and the crossing
can disappear by perturbation. 

The bulk band-structure is gapped for both the QSH and I phases 
and looks similar.
Namely, the topological order is not evident in the bulk band 
structure. The existence of the robust edge states is encoded 
in the bulk wavefunctions. 
There is a bulk-edge correspondence:
\begin{enumerate}
\item $Z_2$ topological number is $\nu=1$ ($\nu=0$).
\item there are odd (even) number of Kramers pairs of gapless edge states. 
\end{enumerate}
This correspondence can be shown 
by the argument similar to the well-known Laughlin's gedanken 
experiment \cite{Laughlin81,Fu06a}.
Suppose we consider the system on a ribbon, with two opposite ends
attached. In one direction the system is periodic while in the 
other directions there are edges. 
When we increase the flux penetrating the hole from zero to half of 
the flux quantum,
the change of a certain physical quantity (time-reversal polarization)
is zero for $\nu=0$ while it is unity for $\nu=1$ \cite{Fu06a}. This means 
that for $\nu=1$ there is a Kramers pair of gapless edge states while for 
$\nu=0$ there are no gapless edge states. 

\subsection{Surface States in 3D Quantum Spin Hall Systems}
The analogous phase is also possible in 3D. 
In this case this phase is an insulator in the bulk and 
supports gapless surface states carrying spin currents.
In this case as well, there is a correspondence between 
the bulk and the surface. The topology of  
 Fermi ``curve'' of the surface states
is related with the $Z_2$ topological numbers for 
the bulk. 

One can see this by the 3D tight-binding model introduced by 
Fu {\it et al.} \cite{Fu06b} on a diamond lattice. 
This model
exhibits a transition between QSH and I phases.
The model is written as 
\begin{equation}
H=t\sum_{\langle ij\rangle}c_{i}^{\dagger}c_{j}
+i(8\lambda_{\mathrm{SO}}/a^{2})\sum_{\langle\langle
ij\rangle\rangle} c_{i}^{\dagger}\mathbf{s}\cdot
(\mathbf{d}_{ij}^{1}\times \mathbf{d}_{ij}^{2})c_{j}.
\end{equation}
Here $a$ is the size of the cubic unit cell, $t$ is the 
nearest-neighbor 
hopping, and ${\bf s}=(s_x, s_y, s_z)$ are the Pauli matrices. 
The term with $\lambda_{\mathrm{SO}}$ is
 a spin-dependent hopping 
to the next nearest neighbor sites, representing the 
spin-orbit coupling.
The vectors  $\mathbf{d}_{ij}^{1}$ and 
$\mathbf{d}_{ij}^{2}$ denote  
those for the
two nearest neighbor bonds involved in 
the next-nearest-neighbor hopping.

In 3D, the TRIMs are ${\bf \Gamma}_{i=(n_1 n_2 n_3)}
=\frac{1}{2}(n_1{\bf G}_{1}+n_2{\bf G}_{2}+n_3{\bf G}_{3})$ 
with $n_1,n_2,n_3=0,1$.
There are four 
$Z_2$ topological numbers 
$\nu_{0},\nu_{1},\nu_{2},\nu_{3}$ \cite{Moore06,Fu06b}, given by
\begin{equation}
(-1)^{\nu_{0}}=\prod_{i=1}^{8}\delta_{i}, \ \
(-1)^{\nu_{k}}=\prod_{n_k=1; n_{j\neq k}=0,1}\delta_{i=(n_1n_2n_3)}.
\label{eq:Z2-3D}
\end{equation}
Each phase is expressed as $\nu_0; (\nu_1\nu_2\nu_3)$, which 
distinguishes 16 phases.
Because among $\nu_i$, $\nu_0$ is the only topological number which 
is robust against disorder, the phases are mainly 
classified by $\nu_0$.
When $\nu_{0}$ is odd the phase is called the strong topological 
insulator (STI), and when $\nu_{0}$ is even it is called the weak topological 
insulator (WTI). The STI and WTI correspond to the QSH and I phases, 
respectively.
The other indices $\nu_1$, $\nu_2$, and $\nu_3$ are used
to distinguish various phases in the STI or WTI phases,
and each phase can be associated with a mod 2 reciprocal lattice vector
$\mathbf{G}_{\nu_1\nu_2\nu_3}=\nu_1\mathbf{b}_{1}
+\nu_2\mathbf{b}_{2}
+\nu_3\mathbf{b}_{3}$, as was proposed in Ref.~\citen{Fu06b}.
These topological numbers in 3D determine the topology of the 
surface states for arbitrary crystal surfaces \cite{Fu06b}.
We note that among the four $Z_2$ topological numbers in 3D, 
only $\nu_{0}$ is robust against nonmagnetic impurities, 
while the others ($\nu_k$ ($k=1,2,3$)) are meaningful only 
for a relatively clean sample \cite{Fu06b}.

\section{Quantum Phase Transition with a Change of 
Topological Number}
As we have seen in the Haldane's model on the honeycomb lattice, 
it is hard to calculate 
the topological number itself and 
to capture its physical meaning in an intuitive way. 
The topological number involves an integral over the whole 
Brillouin zone, both for the QSH systems and the QH systems.

On the other hand, the ``change'' of the topological number
is more accessible. This is because the change occurs locally 
in ${\bf k}$ space. As we change an external parameter,
the system may undergo a phase transition.
It accompanies a closing of a bulk gap at a certain ${\bf k}$, 
because it
is the only way to change the topological number. 
When an external parameter is changed, 
in some cases the phase transition occurs, while in other cases
it does not, because of the level repulsion. 

In the following we consider ``generic'' gap closing by 
tuning a {\it single} parameter, which we call $m$.
We exclude the cases where the gap closing is 
achieved by tuning more than one parameters.
In such cases, the phase transition might be 
circumvented by some perturbation.
In general, energy levels repel each other, thereby the 
valence and the conduction bands do not touch if the number of tuned 
parameters is not large enough. The number of tuned parameters to 
achieve degeneracy, called the codimension, is sensitive to the symmetry 
and the dimension of the system considered.

We henceforth consider only clean systems without any impurities or
disorder. The time-reversal symmetry is
assumed.
We also assume that the Hamiltonian is generic, and 
we exclude the Hamiltonians which require fine tuning of parameters. 
In other words, we exclude the cases which are vanishingly improbable
as a real material.

\subsection{2D Quantum Hall Systems}
For the 2D QH systems, it is simple, and in fact we have already 
seen the physics of gap closing in the Haldane's model. 
We can argue  gap closing in generic systems,
and it becomes similar to the one already described in Section 2.
The Chern number is the total flux of $B_z({\bf k})$ 
inside the Brillouin zone. 
Therefore, if we roll the Brillouin zone into a torus, 
the Chern number is nothing but a total monopole charge
inside the torus, where a monopole and an antimonopole
have charges $+1$ and $-1$, respectively.
Each band is associated with the respective Chern number. 
At the gap closing, the Chern number at the lower band
changes by $\pm 1$ and that of the upper band
changes by $\mp 1$. This means that a monopole (or an 
antimonopole) goes out of the Brillouin-zone torus of the upper band,
and into the Brillouin-zone torus of the lower band.

Let $m$ denote the parameter which drives the phase transition.
In the Haldane's model the on-site staggered potential 
$M$ plays the role of the parameter $m$.
The feature of the phase transition 
is further clarified by considering a hyperspace
$(m,\ k_x, k_y)=(k_0,\ k_1,\ k_2)$, 
and characterizing
the gap closing in terms of the gauge field in $m$-${\bf k}$-space 
\cite{Berry84, Volovik}. 
Suppose the gap closes at an 
isolated point $\tilde{\bf k}=(m,\mathbf{k})$.
Then the involved bands, which we call $\alpha$-th and $\beta$-th bands,
have monopoles in $m$-${\bf k}$ space, with their monopole charges
are opposite in sign \cite{Berry84,Volovik}. More precisely,
the gauge field $\mathbf{A}_{\alpha}
(\tilde{\mathbf{k}})$ and the 
corresponding field strength 
$\mathbf{B}_{\alpha}(\tilde{\mathbf{k}})$ for the $\alpha$-th band 
are defined as
\begin{align}
&\mathbf{A}_{\alpha}
(\tilde{\mathbf{k}})=-i\langle \psi_{\alpha}(\tilde{\mathbf{k}})|\nabla_{\tilde{\mathbf{k}}}|
\psi_{\alpha}(\tilde{\mathbf{k}})\rangle,\label{eq:A}\\
&\mathbf{B}_{\alpha}
(\tilde{\mathbf{k}})=\nabla_{\tilde{\mathbf{k}}}\times\mathbf{A}_{\alpha}(\tilde{\mathbf{k}}).
\label{eq:B}
\end{align}
The corresponding monopole density is defined as 
\begin{equation}
\rho_{\alpha}
(\tilde{\mathbf{k}})=\frac{1}{2\pi}\nabla_{\tilde{\mathbf{k}}}\cdot\mathbf{B}_{\alpha}
(\tilde{\mathbf{k}})
\end{equation}
Except for the point where $\alpha$-th band touches with other bands,
the monopole density $\rho_{\alpha}(\tilde{\mathbf{k}})$ vanishes 
identically. At the $\tilde{\bf k}$ point where the $\alpha$-th band 
touches with another band ($\beta$-th band), the wavefunction cannot be written
as a single analytic function around this point,
and the wavefunction is to be written with more than one ``patches''
which are related by gauge transformation \cite{Kohmoto85}, 
as is similar to the vector potential around the Dirac monopole 
in electromagnetism \cite{WuYang}. 
This leads to a $\delta$-function singularity of 
$\rho(\tilde{\mathbf{k}})$ at the band touching; $\rho_{\alpha}(\tilde{\bf k})
\sim -\rho_{\beta}(\tilde{\bf k})\sim 
\pm\delta(\tilde{\bf k}=\tilde{\bf k}_0)$.  
As a result the monopole density is written in general 
as $\rho(\tilde{\mathbf{k}})=\sum_{l}q_{l}
\delta(\tilde{\mathbf{k}}-\tilde{\mathbf{k}}_{l})$, 
where $q_{l}$ is an 
integer  
representing
a monopole charge.
The monopole charge is conserved under a continuous change of the
Hamiltonian. The monopoles indicate the value of $(m,\mathbf{k})$ where
the gap closes.
At the monopole, the Chern number for the band considered 
changes by $\pm 1$.

As an example, in the Haldane's model on the honeycomb lattice,
the gap closes either at $K$ or at $K'$. When the gap closing occurs 
either at $K$ or at $K'$, the Chern number changes by one. 
Meanwhile, 
when $M$ is changed while $\phi=0$ is kept, 
the gap closes simultaneously at $K$ and $K'$, and the changes of the 
Chern number at $K$ and $K'$ cancel each other; the
Chern number does not change as a result.

\subsection{Quantum Spin Hall Phase and Universal Phase Diagram}
Because the QSH phase is roughly analogous to a 
superposition of two QH systems, the phase transition 
can be studied similarly.
The $Z_2$ topological number is preserved as long as the gap 
remains open. Suppose the system goes from 
the insulator to the QSH phase by changing 
some parameter of the system. Then the gap should close somewhere 
in between. Whether or not gap closes as the parameter is varied reflects
the topological properties of the system.
We investigate the criterion 
for the occurence of the phase transition in Refs.~\citen{Murakami07a,Murakami07b,Murakami08a},
as we explain in the following. Through this study we can show that 
the gap-closing physics is equivalent to the
physics of the $Z_2$ topological number.
\begin{figure}
\centerline{\includegraphics[width=12cm]{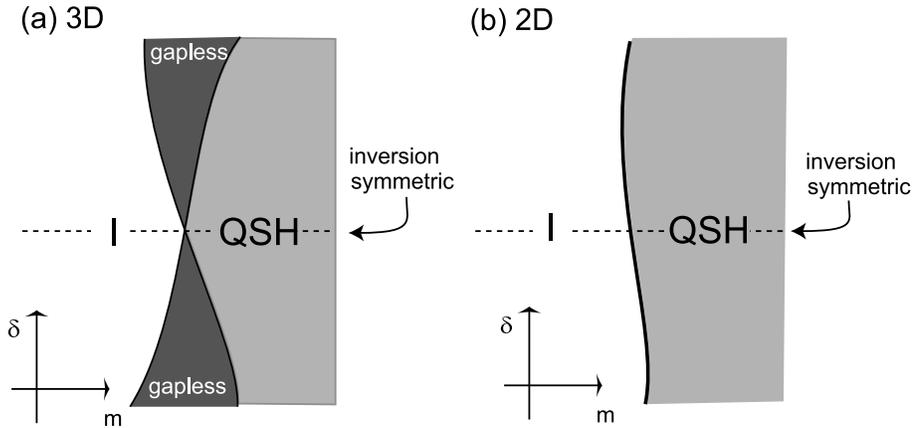}}
\caption{Universal phase diagram between the QSH and the insulator 
phases in (a) 3D and (b) 2D. $m$ is a parameter driving the phase
transition, and $\delta$ is a parameter describing the degree
of inversion-symmetry-breaking.}
\label{fig:phase-diagram}
\end{figure}

\begin{figure}
\includegraphics[scale=0.52]{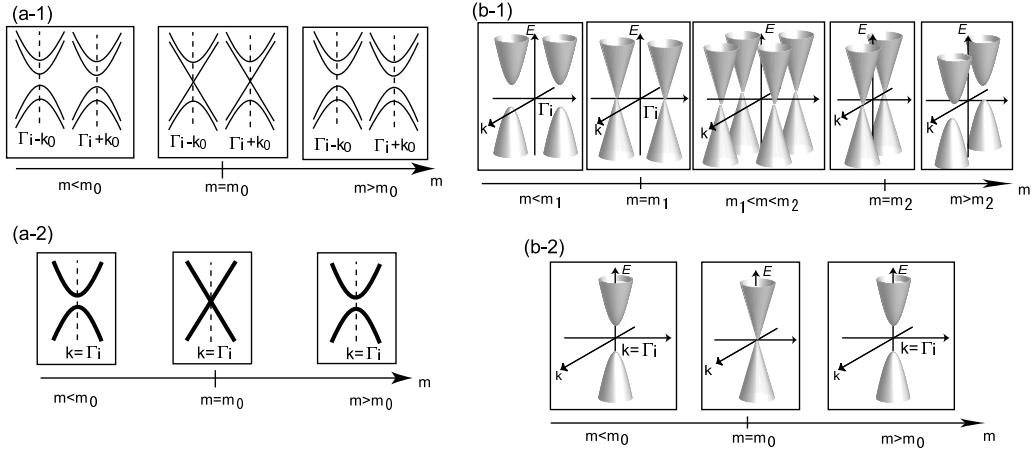}
\caption{Generic gap-closing for (a-1) 2D inversion-asymmetric, 
(a-2) 2D inversion-symmetric, (b-1) 3D inversion-asymmetric and 
(b-2) 3D inversion-symmetric cases. In the cases (a-2)
(b-2) all the states are
doubly degenerate by Kramers theorem. In (a-1)(a-2) and (b-2), the 
gap closing and concomitant phase transition occurs only 
at a single value of $m$: $m=m_0$. Meanwhile in (b-1), by increasing
$m$, 
the gap closes at $m=m_1$, and the bulk remains gapless in 
$m_1\leq m\leq m_2$. The gap opens again at $m=m_2$.
Although in reality the ${\bf k}$ space is two-dimensional in (a-1)(a-2), 
and three-dimensional in (b-1)(b-2), it is drawn as one-dimensional
in (a-1)(a-2), and two-dimensional in (b-1)(b-2) for clarity of illustration.}
\label{fig:degeneracy}\end{figure}

To study the phase
transition in 2D and in 3D, we consider a Hamiltonian matrix
\begin{equation}
H({\bf k})=\left(\begin{array}{cc}
h_{\uparrow\uparrow}({\bf k})& h_{\uparrow\downarrow}({\bf k})\\
h_{\downarrow\uparrow}({\bf k})& h_{\downarrow\downarrow}({\bf k})
\end{array}\right).\label{eq:Hamiltonian}
\end{equation}
We assume that the system is a band insulator, 
and the Fermi energy $E_F$ 
lies within the gap.
The time-reversal-symmetry gives,
\begin{equation}
H({\bf k})=s_y  H^{T}(-{\bf k})s_y,
\label{time-reversal}\end{equation}
which is rewritten as
$h_{\uparrow\uparrow}({\bf k})
=h_{\downarrow\downarrow}^{T}(-{\bf k})$,
$h_{\uparrow\downarrow}({\bf k})
=-h_{\uparrow\downarrow}^{T}(-{\bf k})$,
$h_{\downarrow\uparrow}({\bf k})
=-h_{\downarrow\uparrow}^{T}(-{\bf k})$.
The Kramers theorem guarantees that 
the band structure of such time-reversal-symmetric spin-$1/2$ system is
symmetric with respect to ${\bf k}\leftrightarrow -{\bf k}$.
While the dimension of the Hamiltonian is arbitrary,
it will be taken as the number of states 
involved in the gap closing.

The feature of the phase transition is 
different whether the system considered is (i) ${\cal I}$-symmetric 
or (ii) ${\cal I}$-asymmetric \cite{Murakami07b,Murakami07a}. 
It is because the degeneracy for each state is different
for the two cases.
According to the Kramers theorem, the time-reversal-symmetry says
$\varepsilon_{n \alpha}({\bf k}) = \varepsilon_{n{\bar \alpha}}(-{\bf k})$,
where $\varepsilon_{n \alpha}({\bf k})$ is the energy of 
the $n$-th band with pseudospin $\alpha$, and ${\bar \alpha}$
is the pseudospin opposite to $\alpha$.
If in addition, 
the system is ${\cal I}$-symmetric (i), 
all the states are doubly 
degenerate, because
the ${\cal I}$-symmetry imposes 
$\varepsilon_{n \alpha}({\bf k}) = \varepsilon_{n \alpha}(-{\bf k})$, leading 
to $\varepsilon_{n \alpha}({\bf k}) = \varepsilon_{n {\bar \alpha}}({\bf k})$.
If (ii) ${\cal I}$-symmetry is broken, double degeneracy occurs only at points 
 ${\bf k} = {\bf \Gamma}_{i}$, where ${\bf \Gamma}_{i}$ 
is one of the TRIM.
By analyzing the respective cases (i)(ii) in 2D and 3D, we 
obtain a universal phase diagram shown in Fig:~\ref{fig:phase-diagram}.
Here $m$ is a parameter driving the phase
transition, and a parameter $\delta$ describes the degree
of inversion-symmetry-breaking.   
The derivation of this universal phase diagram is generic and 
based on topological arguments. Hence it  
does not depend on the details of the system. The parameter $m$
can be any parameter, and in the CdTe/HgTe/CdTe quantum well 
the well thickness $d$ plays the role of $m$ here.

\subsubsection{Inversion asymmetric systems}
In ${\cal I}$-asymmetric systems,  
when ${\bf k} \neq  {\bf \Gamma}_{i}$,
each band is non-degenerate. 
At the gap-closing point, one valence band and one conduction band
become degenerate. In this case a $2\times 2$ Hamiltonian matrix
is sufficient for our purpose; 
\begin{equation}
H=\left(
\begin{array}{cc}
a & c\\
c^{*} & b
\end{array}
\right),
\end{equation}
where
$a$, $b$ are real functions of ${\bf k}$ and $m$, and $c$ is 
a complex function of ${\bf k}$ and $m$. 
A necessary condition for the two eigenvalues to be identical 
consists of three conditions $a=b$, ${\rm Re}c=0$ and ${\rm Im}c=0$,
i.e. the codimension is three \cite{vonNeumann29,Herring37}. 
To put it in a different way, the 2$\times$2  Hamiltonian $H(m,\mathbf{k})$ 
is expanded as
\begin{equation}
H(m,\mathbf{k})=a_0(m,\mathbf{k})+\sum_{i=1}^{3}a_{i}(m,\mathbf{k})
\sigma_i.
\end{equation}
The gap closes when the two eigenvalues are identical, i.e.
when the three conditions
$a_{i}(m,\mathbf{k})$=0 ($i=1,2,3$) are satisfied. This means that the codimension is three.

In 2D, the codimension three is equal to the number of parameters involved, 
that is, $k_x$, $k_y$ and $m$. Thus the gap can close at some ${\bf k}$
when the parameter $m$ is tuned to a critical value.
Near the gap-closing point ${\bf k}={\bf k}_{0}(\neq {\bf \Gamma}_{i})$,
the system's Hamiltonian corresponds to massive Dirac fermion, 
and can be expressed 
as 
\begin{equation}
{\cal H}=(m-m_0)\sigma_z 
+(k_x-k_{0x})\sigma_x+(k_y-k_{0y})\sigma_y
\end{equation}
after unitary and scale transformations. 
The time-reversal-symmetry requires that
the gap closes simultaneously at ${\bf k}_0$ and 
$-{\bf k}_0$ as shown in 
Fig.~\ref{fig:degeneracy} (a-1), and that the masses of the Dirac fermions 
at ${\bf k}=\pm 
{\bf k}_{0}$ have 
opposite signs.
In the Kane-Mele model for the QSH phase \cite{Kane05a,Kane05b}
the gap closes simultaneously 
at the $K,K'$ points, corresponding to the present case.

\begin{figure}
\includegraphics[scale=0.45]{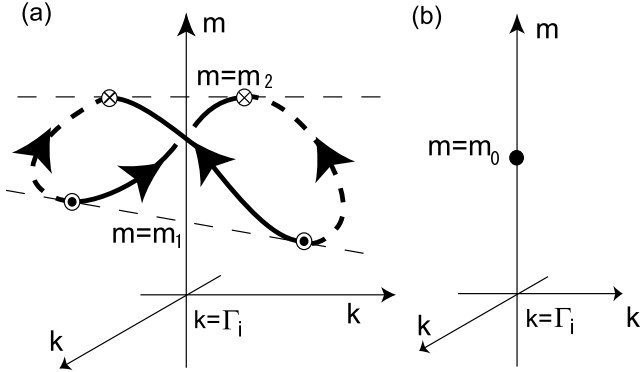}
\caption{Trajectory of the gap-closing points for (a) inversion- 
(${\cal I}$-)asymmetric and 
(b) symmetric systems. For (b) ${\cal I}$-symmetric systems, the gap-closing 
point is located at $\mathbf{k}={{\bf \Gamma}}_i$, and isolated in the
$m$-$\mathbf{k}$ space. Only 
at $m=m_0$ the system is gapless. 
For (a) ${\cal I}$-asymmetric systems, the gap-closing points are created in monopole-antimonopole pairs
at $m=m_1$, and
move in $\mathbf{k}$-space as $m$ changes. 
Solid and broken curves denote the trajectories of the 
monopoles and antimonopoles, respectively. The system opens a gap when 
these gapless points annihilate in pairs at $m=m_2$. }
\label{fig:monopole}\end{figure}

In 3D, as is different from 2D, the gap closing at 
${\bf k}=\pm {\bf k}_{0}\neq {\bf \Gamma}_{i}$ 
cannot lead to phase transion. This is because the 
codimension three is less than the number of 
parameters $(m,k_x,k_y,k_z)$.
The three gap-closing conditions determine a curve 
 in the four-dimensional space $(m,k_x,k_y,k_z)$.
When $m$ is changed continuously the gap-closing ${\bf k}$
point moves in the ${\bf k}$ space, and the system 
remains gapless. 
This curve forms a loop $C$ in $m$-${\bf k}$ space, and 
the gap opens when $m$ is changed across the extremum
of the loop.
The loop $C$ is a trajectory of the gap-closing points
i.e. monopoles
in the $\mathbf{k}$-space as we change $m$. 
It forms a closed loop in the $m$-$\mathbf{k}$ space,
because of the conservation 
of the monopole charge.
The time-reversal-symmetry requires
\begin{equation}
\mathbf{B}_{\alpha}(\mathbf{k})=-\mathbf{B}_{\bar{\alpha}}(-\mathbf{k}), \ 
\rho_{\alpha}(\mathbf{k})=\rho_{\bar{\alpha}}(-\mathbf{k}). 
\end{equation}
The monopoles are symmetric 
with respect to the origin. Therefore, the generic form of the 
loop $C$ is as shown in Fig.~\ref{fig:monopole}.
Thus the loop $C$ occupies a finite region in the value of $m$, and 
it follows that 
in ${\cal I}$-asymmetric 3D systems, gapless phase emerges \cite{Murakami07b}, 
which 
is nonexistent in 2D.

So far we discussed the gap closing at ${\bf k}\neq {\bf \Gamma}_{i}$. 
To complete the discussion, 
we show that the gap does not close at ${\bf k} = {\bf \Gamma}_{i}$,
when the system is ${\cal I}$-asymmetric. 
At ${\bf k}={\bf \Gamma}_{i}$, 
the band
is doubly degenerate, and the codimension 
is five \cite{Avron88,Avron89}, exceeding the number of
tunable parameters which is one (that is, $m$). 
Thus, generic gap-closing
cannot occur at ${\bf k} = {\bf \Gamma}_{i}$.
One can see this as follows.
Because both the valence and the conduction bands 
are doubly degenerate, we consider $4\times 4$ 
Hamiltonian matrix with time-reversal-symmetry.
From (\ref{time-reversal}) we get
\begin{equation}
{H}(m,{\bf k}={\bf \Gamma}_{i})=E_{0}+\sum_{i=1}^{5}a_{i}
\Gamma_{i}
\label{eq:asym-Gamma-i}
\end{equation}
where $a_{i}$'s and $E_{0}$ are real, and
$\Gamma_{1}=1\otimes\tau_{x}$, $\Gamma_{2}=\sigma_{z}\otimes\tau_{y}$, 
$\Gamma_{3}=1\otimes\tau_{z}$, 
$\Gamma_{4}=\sigma_{y}\otimes\tau_{y}$, and  
$\Gamma_{5}=\sigma_{x}\otimes\tau_{y}$, and $\sigma_i$, $\tau_i$ 
are the Pauli matrices.
The eigenenergies are 
$E_{0}\pm\sqrt{\sum_{i=1}^{5}a_{i}^{2}}$. 
The condition for the gap-closing between 
the two (doubly-degenerate) bands consists of five equations
$a_{i}=0$ for $i=1,\cdots,5$, 
which are not satisfied by tuning only one parameter $m$. (Here
the wavenumber ${\bf k}$ is fixed to be ${\bf \Gamma}_{i}$.)
Thus the gap does not close at ${\bf k}={{\bf \Gamma}}_{i}$ 
by tuning a single parameter $m$.

\subsubsection{${\cal I}$-symmetric systems}
In ${\cal I}$-symmetric systems, the energies 
are doubly degenerate for every ${\bf k}$ by the Kramers theorem. 
The gap closes  
between the two doubly-degenerate bands, 
and 
we set the Hamiltonian 
matrix $H({\bf k})$ to be 4$\times$4.
The ${\cal I}$-symmetry is imposed as
\begin{equation}
H(-{\bf k})=PH({\bf k})P^{-1}, \ u(-{\bf k})=Pu({\bf k}),
\label{eq:H-inversion}
\end{equation}
where $P$ is a unitary matrix independent of ${\bf k}$ which 
commutes with the spin matrices, 
and $u({\bf k})$ is the periodic part 
of the Bloch wavefunction: $\varphi_{{\bf k}}({r})
=u({\bf k})e^{i{\bf k}\cdot{r}}$.
Because $P^2=1$, the eigenvalues of $P$ are $\pm 1$. 
By a unitary transformation which diagonalizes $P$, 
one can rewrite
\begin{equation}
P=\left(
\begin{array}{cc}
P_{\uparrow}&\\
&P_{\downarrow}\end{array}
\right),\ \ 
P_{\uparrow}=P_{\downarrow}={\rm diag}(\eta_{a},\ \eta_{b}),\ \ 
\eta_a=\pm 1,\ \ \eta_b=\pm 1
\end{equation}
 without losing generality. $\eta_a$ and $\eta_b$ represent the
parity eigenvalues of the wavefunctions.
One of them corresponds to the valence band, and the other to the 
conduction band.

While there are four combinations for $\eta_a=\pm 1$ and $\eta_b=\pm 1$, 
the overall sign for $(\eta_a,\eta_b)$ is arbitrarily changed 
by gauge transformation. Thus the only distinct case are
(i) $\eta_a=\eta_b$ and 
(ii) $\eta_a=-\eta_b$.
The case (i) $\eta_{a}=\eta_{b}=\pm 1$ means that 
the wavefunctions (orbitals) $a$, $b$ 
have the same parity, e.g. two $s$-like orbitals or two $p$-like 
orbitals.
When $\eta_{a}=\eta_{b}=\pm 1$, the Hamiltonian 
becomes
\begin{equation}
{H}({\bf k})=E_{0}({\bf k})+\sum_{i=1}^{5}a_{i}({\bf k})
\Gamma_{i},
\label{eq:sym-same}
\end{equation}
where $a_{i}$'s and $E_{0}$ are real even functions of ${\bf k}$.
On the other hand, when (ii) $\eta_{a}=-\eta_{b}=\pm 1$, 
where the two constituent wavefunctions have different parity,
the Hamiltonian is
\begin{equation}
{H}({\bf k})=E_{0}({\bf k})+a_{5}({\bf k})\Gamma'_{5}+\sum_{i=1}^{4}
b^{(i)}({\bf k})
\Gamma'_{i}
\label{eq:sym-different}
\end{equation}
where $E_0({\bf k})$ and $a_{5}({\bf k})$ are even functions of ${\bf k}$,
and $b^{(i)}({\bf k})$ are odd functions of ${\bf k}$.
The matrices $\Gamma'_{1}=\sigma_{z}\otimes
\tau_{x}$,
$\Gamma'_{2}=1\otimes\tau_{y}$,
$\Gamma'_{3}=\sigma_{x}\otimes\tau_{x}$,
$\Gamma'_{4}=\sigma_{y}\otimes\tau_{x}$,
and $\Gamma'_{5}=1\otimes\tau_{z}$ form the Clifford algebra.
Therefore, in both cases, $\eta_a=\eta_b$ and $\eta_a=-\eta_b$, 
the codimension is five, which exceeds the number of tunable 
parameters ($m$, $k_x$, $k_y$, $k_z$).
Therefore, gap closing does not occur in general.

However, this counting holds true only when ${\bf k}$ is at a
generic point with ${\bf k}\neq  {{\bf \Gamma}}_{i}$, 
At ${\bf k}={\bf \Gamma}_{i}$,
the codimension (number of parameters to
achieve degeneracy) 
remains 5 for $\eta_a=\eta_b$ (Eq.~(\ref{eq:sym-same})), 
while it becomes 1 for $\eta_a=-\eta_b$ (Eq.~(\ref{eq:sym-different})).
This is because the odd functions $b^{(i)}({\bf k})$ vanish identically
at ${\bf k}={\bf \Gamma}_{i}$, 
and one has only to tune
$a_{5}({\bf k})$ to be zero. 
The number of parameters ($m$) is one, because the wavenumber is fixed
as ${\bf k}={{\bf \Gamma}}_{i}$. Thus, the gap closes by fine-tuning a 
single parameter, only
when $\eta_a=-\eta_b$.

In the following we derive an effective Hamiltonian describing
the low-energy physics of the QSH-I phase transition.
As a gap-closing point, we take ${\bf k}=0$ as an example, and
write down the Hamiltonian explicitly. Extension to other ${\bf k}=
{\bf \Gamma}_{i}$ points is straightforward.
The Hamiltonian is expanded in terms of ${\bf k}$ as
\begin{equation}
{H}(m,{\bf k})\sim E_{0}+m\Gamma'_{5}+\sum_{i=1}^{4}
\left({\bf \beta}^{(i)}\cdot{\bf k}\right)
\Gamma'_{i},
\end{equation}
where $E_0$ and $m$ are constants, and 
${\bf \beta}^{(i)}$ $(i=1,\cdots,4)$ are two-dimensional real constant vectors.
The critical value of $m$ is set as zero. After unitary transformations, 
the Hamiltonian finally becomes block-diagonal,
\begin{equation}
{H}(m,{\bf k})=E_{0}+\left(
\begin{array}{cccc}
m&z_{-}&&\\
z_{+}&-m&&\\
&&m&-z_{+}\\
&&-z_{-}&-m
\end{array}\right).\label{eq:case-b}\end{equation}
where $z_{\pm}=b_{1}k_{x}+b_{3}k_{y}\pm ib_{2}k_{y}$ with real constants
$b_1$, $b_2$ and $b_3$.
If the system has fourfold rotational symmetry in the $xy$ plane for example,
one has $b_{1}=b_{2}$ and $b_{3}=0$, leading to $z_{\pm}\propto 
k_{x}\pm ik_{y}$.
Thus we have shown that
a generic Hamiltonian with time-reversal- and ${\cal I}$- symmetries 
decouples into a pair of 
Hamiltonians describing two-component Dirac fermions,
with opposite signs of the 
mass terms.
Such decoupling is nontrivial. This Hamiltonian
is identical with the one  suggested for the HgTe quantum well 
in Ref.~\citen{Bernevig06f}.
The eigenenergies are
$E=E_{0}\pm \sqrt{m^{2}+z_{+}z_{-}}$ and the gap closes at ${\bf k}=0$ when 
the parameter $m$ is tuned to zero.
This kind of effective model can be used for studying disorder 
effects in the QSH phase \cite{Shindou08}.

\subsubsection{$Z_2$ Topological Number}
We have discussed generic types of gap closing
in time-reversal invariant 
systems, achieved by tuning a single parameter. 
There are two types of gap closing: (a) 
simultaneous gap closing at ${\bf k}=\pm{\bf k}_{0} (\neq{\bf \Gamma}_{i})$ occurs
in systems without ${\cal I}$-symmetry, and (b) gap closing between
two Kramers-degenerate bands (i.e. four bands) at ${\bf k}={\bf \Gamma}_{i}$ occurs in systems with 
${\cal I}$-symmetry 
(see Fig.~\ref{fig:degeneracy}). 
Thus the gap-closing by tuning a single parameter
occurs only in limited cases,
due to level repulsion.
We note that  we have not made any assumption on the $Z_2$ topological number,
in deriving this result.
Nevertheless, we can show that the both cases  
 (a) and (b) involve a change of the $Z_2$ topolgical number,
and are accompanied by a
quantum phase transition.
The Kane-Mele model on the honeycomb lattice \cite{Kane05b} belongs 
to class (a) while the HgTe quantum-well model \cite{Bernevig06f} belongs to class (b).

In 2D ${\cal I}$-symmetric systems, the gap closing at 
the QSH-I 
transition occurs at TRIM $\mathbf{k}={{\bf \Gamma}}_i$. This is
accompanied by an exchange of 
the parity eigenvalues between the valence and 
the conduction bands. It corresponds 
to the expression of the $Z_2$ topological number as a product of 
the parity eigenvalues over all the TRIMs
$\mathbf{k}={{\bf \Gamma}}_i$  over the
occupied states \cite{Fu06a} (Eq.~(\ref{eq:Isym})).
On the other hand, for ${\cal I}$-asymmetric 2D systems,
the gap closes at $\pm 
\mathbf{k}_{0} (\neq {\bf \Gamma}_{i})$ by tuning $m$.
Because the $Z_2$ topological number should change at the gap closing, 
the $Z_2$ topological number
should be expressed as an integral over the $\mathbf{k}$ space. 
This is the Pfaffian expression of the $Z_2$ topological 
number (Eq.~(\ref{eq:Iasym})) \cite{Kane05a,Fu06a}.
In retrospect, the gap closing in the ${\cal I}$-symmetric systems 
occur only at ${\bf \Gamma}_{i}$ between the bands with the opposite parities.
This is because otherwise the codimension is five, exceeding the number 
of tunable parameters.
On the other hand, 
if ${\cal I}$-symmetry is absent, and the level repulsion is less stringent,
and  the gap can close at some ${\bf k}$ other than the TRIM 
${\bf \Gamma}_{i}$. 
This difference is reflected in the definition of the $Z_2$ topological number.

\subsection{Example: 3D Fu-Kane-Mele model}
\begin{figure}
\includegraphics[scale=0.5]{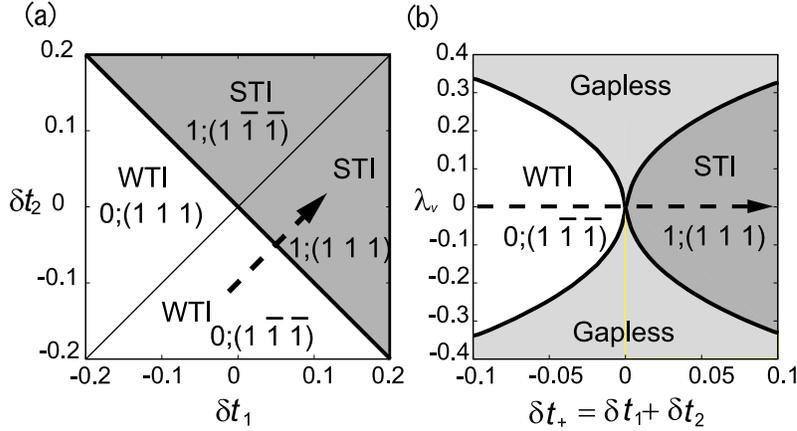}
\caption{Phase diagrams for the Fu-Kane-Mele model with $\delta t_3=0$,
$\delta t_4=0$. $t_1$ and $t_2$ are the bonds along the 111 and 
$1\bar{1}\bar{1}$ directions. 
We put $\lambda_{\mathrm{SO}}=0.1t$. The axes are in the unit
of $t$. (a) The phase diagram in 
$\delta t_1$-$\delta t_2$ plane \cite{Fu06b}.
$\lambda_v$ is set as zero.
(b) The phase 
diagram in the $\delta t_{+}$-$\lambda_v$ plane.
Here $\delta t_{+}=\delta t_{1}+\delta t_{2}$,
while $\delta t_{-}=\delta t_{1}-\delta t_{2}=0.1t$ is fixed. 
The arrows in (a) and (b) refer to the same variation of parameters.}
\label{fig:phase-dia1}
\end{figure}

The 3D Fu-Kane-Mele model \cite{Fu06b} 
is an ideal model for studying 
QSH-I phase transitions in 3D. 
It is a 4-band tight-binding model on a diamond lattice
and is ${\cal I}$- and time-reversal-symmetric. 
It means that every eigenstate is doubly degenerate by the Kramers theorem.
The doubly-degenerate conduction band and the valence band
touch at the three $X$ points, $X^r=(2\pi/a) \hat{r}$ $(r=x,y,z)$.
To describe the phases having a bulk gap,
one sets the nearest-neighbor hoppings for four bond directions to be
different; $t_i$ ($i=1,2,3,4$) \cite{Fu06b}.
The system then opens a gap.
When we set the hopping to be $t_{i}=t+\delta t_{i}$ and 
$\delta t_{3}=0=\delta t_{4}$, the phase diagram is as shown 
in Fig.~\ref{fig:phase-dia1}(a) as a function of $\delta t_{1}$ and 
$\delta t_{2}$, obtained 
in Ref.~\citen{Fu06b}. 
At the phase boundaries the bulk gap vanishes.

To verify the universal phase diagram in 3D, we introduce
the ${\cal I}$-symmetry-breaking term, which 
does not exist in the original Fu-Kane-Mele model. 
The simplest way to break ${\cal I}$-symmetry is
to introduce an alternating on-site energy $\lambda_v$ into
the system, as was done the 2D Kane-Mele model on the
honeycomb lattice \cite{Kane05a}.

We then 
calculate how the WTI-STI phase transition is modified by 
the $\lambda_v$ term.
As we see from the phase diagram (Fig.~\ref{fig:phase-dia1}(a)) ,
we regard $\delta t_{+}=\delta t_{1}+\delta t_{2}$ as a parameter
$m$ driving the phase transition, while $\delta t_{1}-\delta t_{2}$ is
fixed to be $\delta t_{1}-\delta t_{2}=0.1t$ as an example. 
This corresponds to
the  arrow in Fig.~\ref{fig:phase-dia1}(a).
The phase diagram in the $\delta t_{+}$-$\lambda_{v}$ plane 
is calculated as shown in Fig.~\ref{fig:phase-dia1}(b). 
When the ${\cal I}$-symmetry is broken ($\lambda_{v}\neq 0$),
the gapless region appears in the phase diagram, in 
accordance with our universal phase diagram \cite{Murakami08a}.

\begin{figure}[htb]
\includegraphics[scale=0.5]{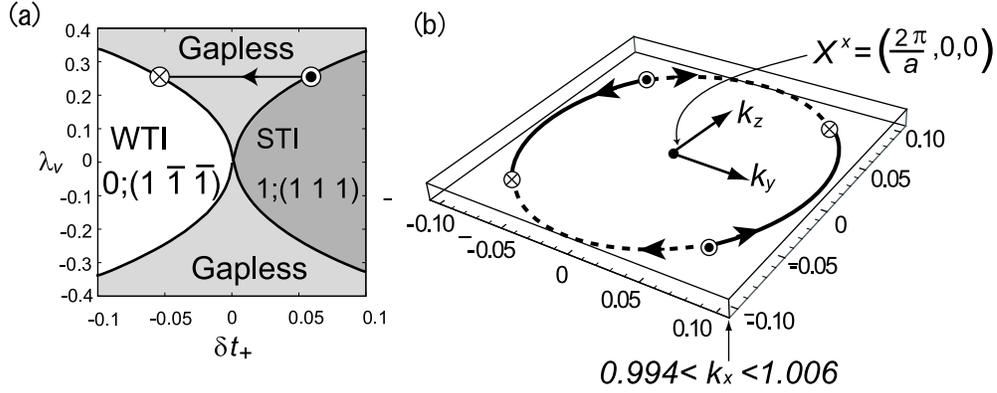}
\caption{Trajectory of the gapless points in $\mathbf{k}$ space.
As we change $\delta t_{+}$ with $\lambda_v$ fixed as shown in 
the arrow in (a), 
the monopoles and antimonopoles
travel in the ${\bf k}$ space as shown in (b) 
by the solid and broken curves, respectively.
The wavenumber $\mathbf{k}$ is shown in the unit of $(2\pi/a)$.}
\label{fig:trajectory-mono2}
\end{figure}

The trajectory (``string'') 
of the gapless points in $\mathbf{k}$ space also agrees with 
our theory.
As the parameter $\delta t_{+}$ is
changed along the arrow in Fig.~\ref{fig:trajectory-mono2}(a), the
gapless points move in $\mathbf{k}$ space as in 
Fig.~\ref{fig:trajectory-mono2}(b) \cite{Murakami08a}. 
The overall feature 
of the trajectory, i.e. its pair creation and annihilation
with changing partners, perfectly agrees with 
our theory.
The change of the $Z_2$ topological numbers is 
also consistent with our theory \cite{Murakami08a}.

\section{Helical edge states}
From the effective model thus obtained, 
we can calculate the helical edge states appearing at the 
boundary between the QSH and the insulator phase. 
Such boundary is described by
setting the mass parameter $m(x)$ to be dependent on space; 
 $m(\pm \infty)= \pm m_{0}$, i.e., 
\begin{equation}
m=\left\{\begin{array}{l}
m_{0}\ \ : x\gg 0\\
-m_{0}\ \ : x\ll 0.
\end{array}\right.
\label{eq:DW}
\end{equation}
The detail of the crossover between $m_{0}$ and
$-m_{0}$ is unimportant and is left unspecified.
For 2D ${\cal I}$-asymmetric systems (Fig.~\ref{fig:degeneracy}(a-1)), one can 
consider the Dirac fermions at ${\bf k}=\pm {\bf k}_{0}$ separately.
Masses of these Dirac fermions change sign at $m=0$; hence they yield 
the edge states localized at the boundary, as explained in 
Ref.~\citen{Niemi86}.
Because the Dirac fermions at ${\bf k}=\pm {\bf k}_{0}$ are 
related  by time-reversal-reversal symmetry, 
the two edge states form a Kramers pair and carry a spin current.

For 2D ${\cal I}$-symmetric systems (Fig.~\ref{fig:degeneracy}(a-2)), we follow the discussion in 
Refs.~\citen{Niemi86,Su79} to 
show that such a boundary between phases with different $Z_2$ 
topological numbers has a Kramers pair of edge states.
By replacing $k_x$ by $-i\partial_x$ in Eq.~(\ref{eq:case-b}), 
we consider 
\begin{eqnarray}
&&\tilde{H}(k_{y})=E_{0}+b_1\partial_x\left(
\begin{array}{cccc}
0&-i &&\\
-i&0&&\\
&&0&i\\
&&i&0
\end{array}\right)\nonumber\\
&&\ \ \ 
+\left(
\begin{array}{cccc}
m& (b_3 -ib_2)k_y&&\\
(b_3 +ib_2)k_y&-m&&\\
&&m&-(b_3 +ib_2)k_y\\
&&-(b_3 -ib_2)k_y&-m
\end{array}\right).
\label{eq:case-b-DM}
\end{eqnarray}
To calculate the eigenstates it is convenient to perform unitary
transformation as
\begin{equation}
H'(k_y)=Q^{\dagger}\tilde{H}({\bf k})Q=
E_{0}+\left(
\begin{array}{cccc}
b_2 k_y& m-b_1\partial_x&&\\
m+b_1\partial_x &-b_2 k_y&&\\
&&-b_2 k_y& m-b_1\partial_x\\
&&m+b_1\partial_x&b_2 k_y
\end{array}\right),\label{eq:case-b-DM-u}\end{equation}
where 
\begin{equation}
Q=\frac{1}{\sqrt{2}}e^{-ib_3 k_{y}x/b_{1}}\left(
\begin{array}{cccc}
1&1&&\\
i&-i&&\\
&&-i&-i\\
&&-1&1
\end{array}\right).
\end{equation}
The eigenvalue problem reads as $H'(k_y)u_{k_{y}}(x)=E(k_y)u_{k_y}(x)$.
The term $E_0$ is absorbed by shifting the energy.
Because (\ref{eq:case-b-DM-u}) is block-diagonal, we first solve 
the eigenvalue problem for the first two components of $u_{k_{y}}$. 
By putting $u_{k_{y}}=(u_1,\ u_2,\ 0,\ 0)^{t}$, we get
\begin{eqnarray}
&&(E-b_2 k_y)u_1=Du_{2},\label{eq:Du2} \\
&&(E+b_2 k_y)u_2=D^{\dagger}u_{1}, \label{eq:Du1} 
\end{eqnarray}
where $D=m-b_{1}\frac{\partial}{\partial x}$, 
$D^{\dagger}=m+b_{1}\frac{\partial}{\partial x}$.
They yield eigenequations for $u_{1}$ and $u_{2}$, respectively:
\begin{eqnarray}
&&DD^{\dagger}u_{1}=(E^{2}-b_{2}^{2}k_{y}^{2})u_{1},\label{eq:DDdag}\\
&&D^{\dagger}Du_{2}=(E^{2}-b_{2}^{2}k_{y}^{2})u_{2}.\label{eq:DdagD}
\end{eqnarray}
Because (\ref{eq:DDdag}) is invariant under
$E\rightarrow -E$, the resulting spectrum  
seems to be symmetric with respect to $E=0$; $E\leftrightarrow -E$. 
However, it is not true, because in some cases the $u_1$ solutions to 
(\ref{eq:DDdag}) has no corresponding solution for $u_2$. 
If $E=-b_2 k_y$, (\ref{eq:Du1})
cannot be solved for $u_{2}$. Similarly, if  
$E=b_2 k_y$, (\ref{eq:Du1}) cannot be solved for $u_1$. Thus 
the solutions which are not symmetric with respect to $E=0$ are
as follows.
For $u_{1}(\neq 0)$ which satisfies $D^{\dagger}u_{1}=0$,
we get $E=b_{2}k_{y}$ and $u_2=0$ from Eqs.~(\ref{eq:Du2}) and
(\ref{eq:Du1}), whereas there is no 
solution with $E=-b_{2}k_{y}$. In the same token, for 
$u_{2}$ which satisfies $Du_{2}=0$, we get 
 $E=-b_{2}k_{y}$ from (\ref{eq:Du2}), whereas there is no 
solution with $E=b_{2}k_{y}$. 
Hence the spectral asymmetry is related to 
the kernels for $D$ and $D^{\dagger}$.
For example, for $b_1>0$ and $m_0>0$, the solution at the 
boundary  (\ref{eq:DW}), 
with $D^{\dagger}u_{1}=0$ gives 
\begin{equation}
u_{1}\propto
\exp\left(-b_{1}^{-1}\int^{x}m(s)ds\right)
\end{equation} and $E=b_{2}k_{y}$, 
while $Du_{2}=0$ has no normalizable solution. 
Thus the energy dispersion in $k_{y}$ direction has a branch $E=b_{2}k_{y}$,
which crosses the Fermi energy $E\sim 0$. This state is gapless, 
localized near $x=0$.
\begin{figure}[h]
\includegraphics[scale=0.9]{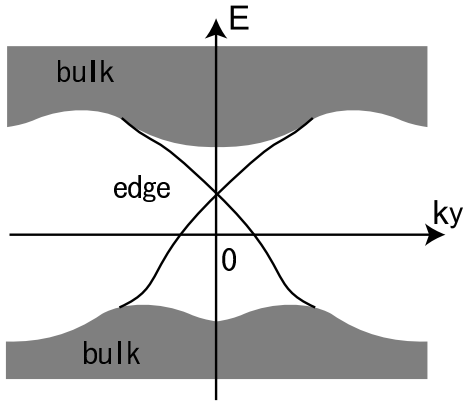}
\caption{Schematic dispersion curves for 
the model (\ref{eq:case-b-DM}).}
\label{fig:dispersion}\end{figure}
We have thus far solved the eigenequation for the first two components.
The lower two components of the wavefunction $u$ is obtained from above by 
time-reversal operation.
Therefore, the above-mentioned edge state with $E=b_2 k_y$ has a Kramers
partner with $E=-b_{2} k_{y}$. The whole dispersion is shown in 
Fig.~\ref{fig:dispersion}.
Thus we have shown that 
the Kramers pair of edge states exists 
at the boundary between the QSH and I phases.
They cross at $k_y=0$, as follows from the Kramers theorem.

\section{Bismuth Ultrathin Films}
The QSH phase requires no magnetic field. This means that some 
materials might realize the QSH by themselves without applying
any field. The only necessary conditions for the QSH systems
are as follows.
\begin{enumerate}
\item nonmagnetic insulator
\item the $Z_2$ topological number is odd ($\nu=1$)
\end{enumerate}
The latter condition means that the spin-orbit coupling 
should be strong enough, which requires relatively heavier elements.  
In the absence of the spin-orbit coupling the $Z_2$ topological 
number $\nu$ is zero (i.e. even). When the spin-orbit coupling
is made stronger, some systems can change its $Z_2$ topological number
from $\nu=0$ to $\nu=1$. At the phase transition, the gap closes. 

In this sense, the gap should be opened by the spin-orbit coupling. 
This is an ambiguous statement. In Ref.~\citen{Murakami06a} we
clarified that systems with large 
susceptibility is a good starting point for the 
materials search.
Among materials with heavy elements, we pick up bismuth as a candidate.
Bismuth is known as a strong diamagnet, due to interband 
matrix elements by the spin-orbit coupling. 
In this sense the gap is originated from the spin-orbit coupling, 
and is a good candidate for the 2D QSH. 
Bismuth itself is a nonmagnetic semimetal, not an insulator. 
We have to open a gap by some means to make it the QSH phase.
One idea is to make it into thin film. Indeed, the recent experiments
and first-principle calculations show that the (111) 1-bilayer 
bismuth has indeed a gap \cite{Koroteev08}. 
In 
Ref.~\citen{Murakami06a} we considered the (111) 1-bilayer bismuth ultrathin 
film from the 
simple tight-binding model, and theoretically proposed that it is the 
QSH phase.
We also calculate the parity eigenvalues for the tight-binding 
model, and confirmed the result. 

The other way to open a gap is to make an alloy with Sb. This leads us to
the 3D QSH, as has been proposed in Ref.~\citen{Fu06b}. 
In the ARPES measurement \cite{Hsieh}, the Fermi ``surface'' for
the surface states has been observed, and it crosses the Fermi
energy odd times, from the $\bar{\Gamma}$ point to the $\bar{M}$ point.
This shows that the surface state of Bi$_{0.9}$Sb$_{0.1}$ is
the topological one for the QSH system.

\section{Concluding Remarks}
With simple examples we have seen various kinds of edge states.
In graphene, the existence of edge states is sensitive to boundary 
conditions, while in the QH and the QSH systems, the gapless edge states
exist irrespective of the boundary conditions. 
This comes from the nontrivial topological number carried by 
the bulk states, defined only for insulators. 
On the other hand, the graphene, being a zero-gap semiconductor,
cannot have such topological numbers, 
which means that the edge states are not robust against perturbations.

The quantum phase transition between the QSH and insulator phases
is studied. We consider generic time-reversal-invariant system
with a gap, and study the condition when the bulk gap closes
by tuning a single external parameter. Due to level repulsion, 
the gap does not always close by tuning a single parameter;
instead in many cases, fine tuning of more than one parameters is
needed to close the gap.
In the ${\cal I}$-symmetric systems, the gap closes only at
the TRIM ${\bf k}={\bf \Gamma}_{i}$, between the valence 
and conduction bands with opposite parities.
In the ${\cal I}$-asymmetric systems, on the other hand, 
the phase transition 
is different between 2D and 3D. In 2D the gap closes simultaneously at 
${\bf k}=\pm {\bf k}_{0}\neq {\bf \Gamma}_{i}$. In 3D there appears a
gapless region in the phase diagram between the QSH and the insulator 
phases. The gap closing points are monopoles and antimonopoles, and
they are created and annihilated in pairs, when the system 
transits from the gapless phase into the phases with a bulk gap (i.e. 
QSH or insulator phases).

It is interesting to note that in each case with robust
gapless edge states, there is an associated current.
In the 2D QH phase it is the charge current which the edge states carry.
In the 2D QSH phase it is the spin current. 
There are also other classies of systems which show this kind
of stable edge states.
One is the superconductor without time-reversal symmetry,
for example with gap function $p_x +ip_y$. In this 
case the current of Majorana fermion is carried by the 
edge states, and the edge states are nothing but the 
surface Andreev bound states. 
Another case of interest is found among 
the superconductors/superfluids
with time-reversal symmetry. 
In this case the edge carries a spin current of the Majorana fermion
\cite{Qi08}.

\section*{Acknowledgements}
We are grateful to N. Nagaosa, S.-C. Zhang, S. Iso, Y. Avishai, M. Onoda
R. Shindou and S. Kuga 
for collaborations and fruitful discussions, and 
R.~Tsukui for making some of the 
figures for the present paper.
Part of this work is based on discussions 
during Yukawa International Seminar 2007 (YKIS 2007)  
entitled as ``Interaction and Nanostructural Effects in 
Low-Dimensional Systems''. 
This research is partly supported in part 
by Grant-in-Aids  
from the Ministry of Education,
Culture, Sports, Science and Technology of Japan.

%

\end{document}